\documentclass[11pt]{article}
\usepackage{graphicx}
\usepackage{epsfig}
\usepackage{epsf}
 
\RequirePackage{xspace}

\usepackage{relsize}
\def\babar{\mbox{\slshape B\kern-0.1em{\smaller A}\kern-0.1em
    B\kern-0.1em{\smaller A\kern-0.2em R}}}

\def\s     {\ensuremath{s}\xspace}

\def\b     {\ensuremath{b}\xspace}

\def\piz   {\ensuremath{\pi^0}\xspace}

\def\pip   {\ensuremath{\pi^+}\xspace}
\def\pim   {\ensuremath{\pi^-}\xspace}

\def\pipm  {\ensuremath{\pi^\pm}\xspace}

\def\Kbar  {\kern 0.2em\overline{\kern -0.2em K}{}\xspace}

\def\Kz    {\ensuremath{K^0}\xspace}
\def\Kzb   {\ensuremath{\Kbar^0}\xspace}
\def\KzKzb {\ensuremath{\Kz \kern -0.16em \Kzb}\xspace}
\def\Kp    {\ensuremath{K^+}\xspace}
\def\Km    {\ensuremath{K^-}\xspace}

\def\KpKm  {\ensuremath{\Kp \kern -0.16em \Km}\xspace}
\def\KS    {\ensuremath{K^0_{\scriptscriptstyle S}}\xspace} 
\def\KL    {\ensuremath{K^0_{\scriptscriptstyle L}}\xspace}

\def\Dbar    {\kern 0.2em\overline{\kern -0.2em D}{}\xspace}

\def\Dz      {\ensuremath{D^0}\xspace}
\def\Dzb     {\ensuremath{\Dbar^0}\xspace}
\def\DzDzb   {\ensuremath{\Dz {\kern -0.16em \Dzb}}\xspace}
\def\Dp      {\ensuremath{D^+}\xspace}
\def\Dm      {\ensuremath{D^-}\xspace}

\def\DpDm    {\ensuremath{\Dp {\kern -0.16em \Dm}}\xspace}
\def\Dstar   {\ensuremath{D^*}\xspace}

\def\Dstarmp {\ensuremath{D^{*\mp}}\xspace}

\def\B       {\ensuremath{B}\xspace}
\def\Bbar    {\kern 0.18em\overline{\kern -0.18em B}{}\xspace}

\def\BB      {\ensuremath{B\Bbar}\xspace} 
\def\Bz      {\ensuremath{B^0}\xspace}
\def\Bzb     {\ensuremath{\Bbar^0}\xspace}
\def\BzBzb   {\ensuremath{\Bz {\kern -0.16em \Bzb}}\xspace}
\def\Bu      {\ensuremath{B^+}\xspace}
\def\Bub     {\ensuremath{B^-}\xspace}
\def\Bp      {\ensuremath{\Bu}\xspace}
\def\Bm      {\ensuremath{\Bub}\xspace}

\def\BpBm    {\ensuremath{\Bu {\kern -0.16em \Bub}}\xspace}

\def\BorBbar    {\kern 0.18em\optbar{\kern -0.18em B}{}\xspace}
\def\DorDbar    {\kern 0.18em\optbar{\kern -0.18em D}{}\xspace}
\def\KorKbar    {\kern 0.18em\optbar{\kern -0.18em K}{}\xspace}

\mathchardef\Upsilon="7107
\def\Y#1S{\ensuremath{\Upsilon{(#1S)}}\xspace}

\def\FourS {\Y4S}

\mathchardef\Deltares="7101
\mathchardef\Xi="7104
\mathchardef\Lambda="7103
\mathchardef\Sigma="7106
\mathchardef\Omega="710A

\def\Deltabar{\kern 0.25em\overline{\kern -0.25em \Deltares}{}\xspace}
\def\Lbar{\kern 0.2em\overline{\kern -0.2em\Lambda\kern 0.05em}\kern-0.05em{}\xspace}
\def\Sigbar{\kern 0.2em\overline{\kern -0.2em \Sigma}{}\xspace}
\def\Xibar{\kern 0.2em\overline{\kern -0.2em \Xi}{}\xspace}
\def\Obar{\kern 0.2em\overline{\kern -0.2em \Omega}{}\xspace}
\def\Nbar{\kern 0.2em\overline{\kern -0.2em N}{}\xspace}
\def\Xb{\kern 0.2em\overline{\kern -0.2em X}{}\xspace}

\def\BR         {{\ensuremath{\cal B}\xspace}}

\newcommand{\tev}{\ensuremath{\mathrm{\,Te\kern -0.1em V}}\xspace}
\newcommand{\gev}{\ensuremath{\mathrm{\,Ge\kern -0.1em V}}\xspace}
\newcommand{\mev}{\ensuremath{\mathrm{\,Me\kern -0.1em V}}\xspace}
\newcommand{\kev}{\ensuremath{\mathrm{\,ke\kern -0.1em V}}\xspace}
\newcommand{\ev}{\ensuremath{\mathrm{\,e\kern -0.1em V}}\xspace}
\newcommand{\gevc}{\ensuremath{{\mathrm{\,Ge\kern -0.1em V\!/}c}}\xspace}
\newcommand{\mevc}{\ensuremath{{\mathrm{\,Me\kern -0.1em V\!/}c}}\xspace}
\newcommand{\gevcc}{\ensuremath{{\mathrm{\,Ge\kern -0.1em V\!/}c^2}}\xspace}
\newcommand{\mevcc}{\ensuremath{{\mathrm{\,Me\kern -0.1em V\!/}c^2}}\xspace}

\def\cm   {\ensuremath{\rm \,cm}\xspace}

\def\invfb   {\ensuremath{\mbox{\,fb}^{-1}}\xspace}

\def\mus  {\ensuremath{\rm \,\mus}\xspace}

\def\mus        {\ensuremath{\,\mu{\rm s}}\xspace}    

\def\to                 {\ensuremath{\rightarrow}\xspace}

\def\pep2{PEP-II}

\def\gsim{{~\raise.15em\hbox{$>$}\kern-.85em
          \lower.35em\hbox{$\sim$}~}\xspace}
\def\lsim{{~\raise.15em\hbox{$<$}\kern-.85em
          \lower.35em\hbox{$\sim$}~}\xspace}

\def\CP                {\ensuremath{C\!P}\xspace}

\def\deltat{\ensuremath{{\rm \Delta}t}\xspace}
\def\deltamd{\ensuremath{{\rm \Delta}m_d}\xspace}

\xspace

\newcommand{\epjBase}        {Eur.\ Phys.\ Jour.\xspace}
\newcommand{\jprlBase}       {Phys.\ Rev.\ Lett.\xspace}
\newcommand{\jprBase}        {Phys.\ Rev.\xspace}
\newcommand{\jplBase}        {Phys.\ Lett.\xspace}

\newcommand{\epjc}      [1]  {\epjBase\ C~{\bf #1}}

\newcommand{\plb}       [1]  {\jplBase\ B~{\bf #1}}

\newcommand{\jprl}      [1]  {\jprlBase\ {\bf #1}}
\newcommand{\jprd}      [1]  {\jprBase\ D~{\bf #1}}

\newcommand{\progtp}    [1]  {{Prog.\ Th.\ Phys.\ {\bf #1}}}

\def\jetset74   {\mbox{\tt Jetset \hspace{-0.5em}7.\hspace{-0.2em}4}\xspace}

\def\dt{\Delta t}

\def\mmiss{m_{\rm miss}}

\def\btodstpipm{\Bz \rightarrow \Dstarmp\pi^\pm}

\def\r{{r^{(*)}}}

\def\rsq{\r^{2}}

\def\deltaPhase{\delta^{(*)}}

\newcommand{\SLACPubNumber} {10695}
\newcommand{\LANLNumber} {0409023}

\begin{document}

{\pagestyle{empty}

\begin{flushright}
SLAC-PUB-\SLACPubNumber \\
hep-ex/\LANLNumber \\
Sep 2004 \\
\end{flushright}
                                                                                                                                                                                  
\par\vskip 5cm

\begin{center}
\Large \bf
CKM Phase Measurements
\end{center}
\bigskip
                                                                                                                                                                                  
\begin{center}
{Sergey Ganzhur, \\ {\it DSM/Dapnia, CEA/Saclay, F-91191 Gif-sur-Yvette, France}\\ {\it e-mail:}ganzhur@hep.saclay.cea.fr}
\mbox{ }\\
\end{center}
\bigskip \bigskip
                                                                                                                                                                                  
\begin{center}
{\large \bf Abstract}
\end{center}
Recent experimental results on \CP violation in the \B sector from  \babar\ and {\it BELLE}, 
experiments at asymmetric $e^+e^-$ \B-Factories, are summarized in these proceedings.
The constraint on the position of the apex of the unitary triangle, 
obtained from these measurements  allows a test of the CKM interpretation  
of \CP violation in the Standard Model.
\vfill
\begin{center}
Proceedings to the 15th International Topical Conference on Hadron Collider Physics, HCP2004\\
14 June---18 June 2004, East Lansing MI, USA 
\end{center}

\newpage
} 

\section*{Introduction}

The violation of \CP symmetry is a fundamental property of Nature which plays
a key role in the understanding of the evolution of the Universe.
The Cabibbo-Kobayashi-Maskawa (CKM)
quark-mixing matrix~\cite{ref:km} is a source 
of \CP violation in the Standard Model (SM) and is under
experimental investigation aimed over constraining its parameters. A
crucial part of this program is the measurement of the three angles

\begin{eqnarray*}
\alpha (\phi_2) &=&
\arg{\left(- V^{}_{td} V_{tb}^\ast/ V^{}_{ud} V_{ub}^\ast\right)} \\
\beta(\phi_1) &=&
\arg{\left(- V^{}_{cd} V_{cb}^\ast/ V^{}_{td} V_{tb}^\ast\right)} \\
\gamma (\phi_3) &=&
\arg{\left(- V^{}_{ud} V_{ub}^\ast/ V^{}_{cd} V_{cb}^\ast\right)} \\
\end{eqnarray*}
of the unitary triangle (UT), which represents the unitarity of the CKM matrix. These angles can be extracted from the measured time-dependent \CP asymmetry in 
the different neutral \B decay channels.
The independent measurements of $\alpha$, $\beta$ and $\gamma$  allows us to verify the unitary relation ($\alpha+\beta+\gamma=\pi$), 
resolve the several-fold ambiguity on the angles which usually arises  from one single measurement, and 
search for New Physics (NP), comparing magnitudes of the same angle measured with modes dominated by either tree or penguin amplitudes~\cite{ref:PhysBook}.

\begin{figure}
\begin{center}
  \includegraphics[width=0.55\textheight]{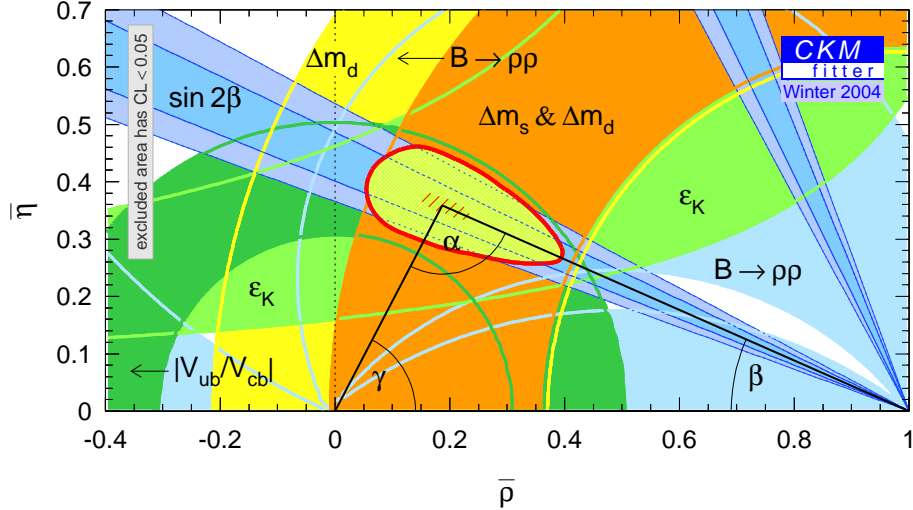}
  \caption{Confidence levels in the complex ($\bar{\rho},\bar{\eta}$) plane obtained from the global fit. The constraint from the world average $\sin(2\beta)/(\phi_1)$ 
is not included in the fit and is overlaid. }
  \label{fig:CKMFit}
\end{center}
\end{figure}

The apex ($\bar{\rho},\bar{\eta}$)~\cite{ref:wolfen} of the UT is already constrained from measurements
which are not involve the \CP violation in the \B meson system. 
From the measured amplitudes of the CKM matrix elements, the mixing frequency of the $\B_d$ and $\B_s$ mesons, and 
the magnitude of indirect \CP violation in the kaon system,  one obtains a 95\% confidence interval for the UT angles~\cite{ref:CKMFitter}:
\begin{eqnarray*}
20.2^\circ < &\beta& < 26.0^\circ \nonumber \\
77^\circ < &\alpha& < 120^\circ \nonumber \\
39^\circ < &\gamma& < 80^\circ \nonumber \\
\end{eqnarray*}
Figure~\ref{fig:CKMFit} shows the constraint in the ($\bar{\rho},\bar{\eta}$) plane obtained from such a fit. 
Thus, the direct measurement of the unitary angles  in \B meson decays will allow us
to check the CKM interpretation of the \CP violation phenomenon in the SM.

\section{Status of the B-Factories}

It is fair to say that most of \CP violation measurements in \B meson decays are coming from $e^+e^-$ energy-asymmetric machines (B-Factories). 
There are  two \B-Factories, PEP-II at SLAC (USA) 
and KEKB at KEK (Japan).  Thanks to a recently incorporated technical feature known as ``trickle'' injection, both achieved luminosities of order 10$^{34}\cm^{-2}\s^{-1}$. 
Two similar asymmetric detectors, \babar~\cite{ref:babar} and {\it BELLE}~\cite{ref:belle} operated at PEP-II and KEKB, respectively, measure charged tracks by a combination
of a silicon vertex detector and a drift chamber embedded in a 1.5~T solenoidal magnetic field. A ring-imaging Cherenkov detector (DIRC) is used for charged
particle identification in \babar\ while {\it BELLE} uses aerogel cherenkov counters (ACC) and a time-of-flight system.  
Both detectors use a CsI(Tl) electromagnetic calorimeter (EMC) to detect photons and identify electrons. The detectors  are also equipped with
muon chambers to identify muons and reconstruct \KL mesons. The key performances of the two experiments are summarized in the following table:
\begin{center}
\begin{tabular}{cccc}\\ \hline
Experiment     & Peak Lum.   &   Best month  & Analyzed  data sample  \\ \hline
\babar         &  $8.8\times 10^{33}\cm^2\s^{-1}$&$15.4\invfb$ &115\invfb (123 M $B\bar{B}$ pairs) \\
{\it BELLE}    &  $13.0\times 10^{33}\cm^2\s^{-1}$&$22.7\invfb$ &140\invfb (152 M $B\bar{B}$ pairs) \\ \hline
\end{tabular}
\end{center}

\section{Experimental aspects}

$e^+e^-$ collisions at the \FourS resonance is a way to produce \BB pairs in a coherent state.
Due to limited phase space, the \B mesons from \FourS are produced almost at rest in the center-of-mass (CM) frame.
That is why the beam energies are different in order to boost the produced \B mesons  with a 
$\beta\gamma=0.56(0.43)$ for \babar\ ({\it BELLE}).
This enables the measurement of the time-dependent \CP asymmetry in the decays of neutral \B mesons.
The method is described in details elsewhere in~\cite{ref:PhysBook}.

The time-dependent \CP\ asymmetry is obtained by measuring the proper time difference \deltat\ between
a fully reconstructed neutral $B$ meson ($\B_{cp}$) decaying into a given final state, and the partially reconstructed
recoil $B$ meson ($\B_{tag}$).  The asymmetry in the decay rate ${\mbox{f}}_+({\mbox{f}}_-)$
when the tagging meson is a \Bz~(\Bzb) is given as
\begin{eqnarray}
{\mbox{f}}_\pm(\, \deltat)& = &{\frac{{\mbox{e}}^{{- \left| \deltat
\right|}/\tau_{\Bz} }}{4\tau_{\Bz}}}  \, [
\ 1 \hbox to 0cm{}
\pm
S \sin{( \deltamd  \deltat )}
\mp
\,C  \cos{( \deltamd  \deltat) }   ],
\label{eq:timedist}
\end{eqnarray}
where $\tau_{\Bz}$ is the \Bz\ lifetime and \deltamd\ is the \Bz--\Bzb mixing frequency.
The parameters $C$ and $S$ describe the magnitude of \CP violation in the decay and
in the interference between decay and mixing (mixing-induced), respectively.
We expect $C=0$ in the case of a single dominant decay amplitude, because the direct \CP\ violation
requires at least two comparable amplitudes with different \CP violating phases, while $S$ is linked to the CKM phases, e.g. $\Bz\to J/\psi\KS$.
Presence of more than one decay amplitudes can lead to  $C\neq 0$ and a non trivial relation of $S$ with unitary angles, e.g. $\Bz\to \pi^+\pi^-$.

\section{CKM phase {\boldmath $\beta/(\phi_1)$}}

\subsection{Charmonium modes}

\begin{figure}
\begin{center}
  \includegraphics[height=.35\textheight]{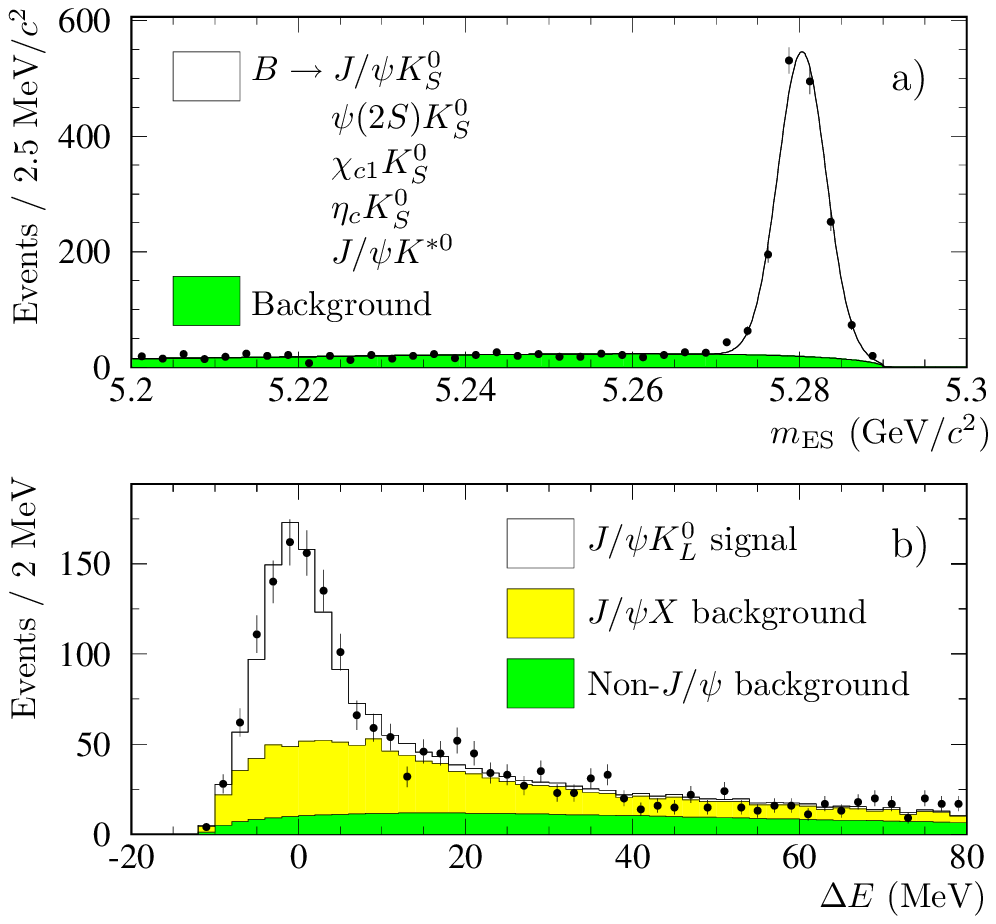}\hfill
  \includegraphics[height=.35\textheight]{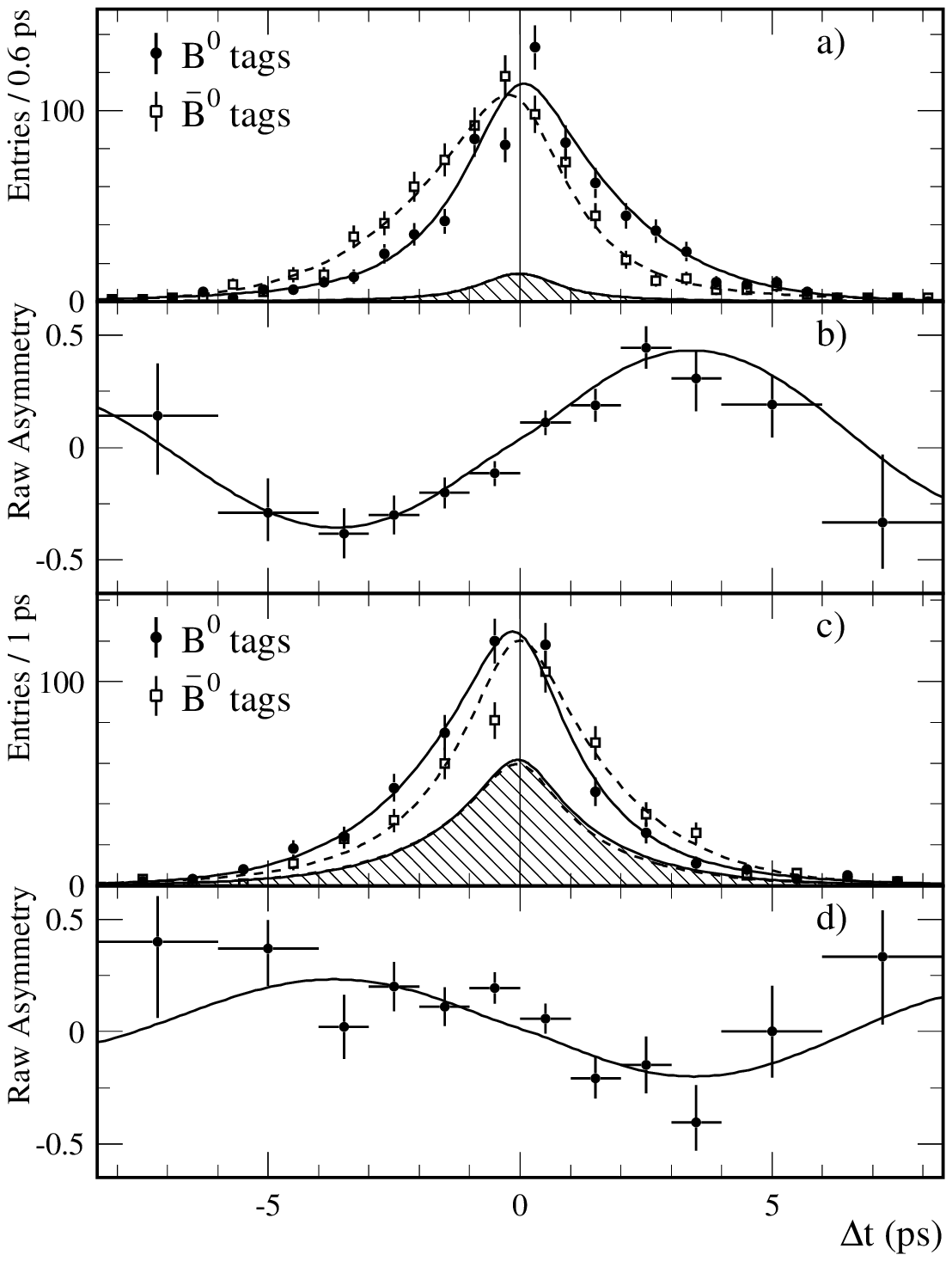}
  \caption{Distributions for $B_{cp}$  candidates: a) beam energy constrained mass for $\eta_f=-1$ decay modes and b) energy difference for $n_f=+1$ $J/\psi\KL$  
(left). Number of \B candidates as function of $\Delta t$ 
for $n_f=-1$ (a) and $J/\psi\KL$ (c) modes (right). The raw asymmetry as function of $\dt$ for $n_f=-1$ (b) and $n_f=+1$ $J/\psi\KL$ (d) modes (right). 
The plots are  from \babar.}
  \label{fig:cc_babar}
\end{center}
\end{figure}

The observation of \CP violation in the \Bz system has been reported in 2000 by \babar\ and {\it BELLE } collaborations. 
New precise measurements of $\sin2\beta$ with a set of charmonium modes similar to the gold plated $J/\psi\KS$ decay channel were reported in
\cite{ref:sin2b_babar,ref:sin2b_belle}. The data sample of 88 (152) millions \BB pairs has been used by \babar\ ({\it BELLE}) to fully reconstruct
a sample of neutral \B mesons decaying into \CP eigenstates such as $J/\psi\KS,$ $\psi(2S)\KS,$ $\chi_{c1}\KS,$ $\eta_{c}\KS$ (\CP-odd)
and $J/\psi\KL$ (\CP-even) as well as vector-vector final state $J/\psi K^*$ which represents a mixture of \CP-even and \CP-odd 
states~\footnote{The angular analysis is required to determine the fraction of \CP-even eigenstate~\cite{ref:LaThuile}}.
The obtained results are summarized in Table~\ref{tab:sin2b}, where the 
two experiments are in good agreement within experimental errors. It is interesting to note that the statistical error is still dominant.  
The average of the two experiments~\cite{ref:HFAG}  
\begin{eqnarray}
\sin2\beta & = & 0.739\pm0.049
\label{eq:sin2bWA}
\end{eqnarray}
is  in a good agreement with Standard Model predictions.  
Figure~\ref{fig:cc_babar} shows  the \B mass (or energy difference) 
and time distributions for \Bz and \Bzb \CP-even and \CP-odd and 
the raw asymmetry $A_{CP}=({\mbox{f}}_+-{\mbox{f}}_-)/({\mbox{f}}_++{\mbox{f}}_-)$ as a function of $\dt$.

\begin{table}
\begin{center}
\begin{tabular}{lcc} \hline
  Mode                                         &           \babar                  & {\it BELLE}  \\
  & ( $88\times 10^6$\BB)            &  ($152\times 10^6$\BB)\\ \hline
  $J\psi\KS(\KS\to\pi^+\pi^-)$    &  $0.82\pm0.08$                    &         $0.67\pm0.08$          \\
  $J\psi\KS(\KS\to\pi^0\pi^0)$    &  $0.39\pm0.24$                    &         $0.72\pm0.20$          \\
  $\psi(2S)\KS(\KS\to\pi^+\pi^-)$ &  $0.69\pm0.24$                    &         $0.89\pm0.20$          \\
  $\chi_{c1}\KS$                  &  $1.01\pm0.40$                    &         $1.54\pm0.49$          \\
  $\eta_{c}\KS$                   &  $0.59\pm0.32$                    &         $1.32\pm0.29$          \\ \hline
All with $\eta_f=-1$                                     &  $0.76\pm0.07$                    &         $0.73\pm0.06$          \\ \hline
  $J\psi\KL$                      &  $0.72\pm0.16$                    &         $0.80\pm0.13$          \\
  $J\psi K^{*0}(K^{*0}\to\KS\pi^0)$    &  $0.22\pm0.52$                    &         $0.10\pm0.45$          \\ \hline 
All charmonium modes   &    $0.74\pm0.07\pm0.03$                    &           $0.73\pm0.06\pm0.03$ 
 \\ \hline
\end{tabular}
\caption{The \CP asymmetry ($\sin2\beta$) measured in the different charmonium decay channels.}
\label{tab:sin2b}
\end{center}
\end{table}

The two vector final state  $J/\psi K^*$ can also be used to measure the sign and magnitude of $\cos2\beta$. Knowledge of the $\cos2\beta$ sign allows us to reduce 
the four-fold ambiguity in the $\beta$ angle. The simultaneous time-dependent and angular analysis for this decay channel obtained by \babar\ where the $\sin2\beta$ 
is fixed to the world average value (\ref{eq:sin2bWA}) favors a positive sign for $\cos2\beta$~\cite{ref:cos2b}:
\[
\cos2\beta = +2.72^{+0.50}_{-0.79}(stat)\pm0.27(syst)
\]

\subsection{Penguin dominated modes}

\begin{figure}
\begin{center}
  \includegraphics[width=0.5\textheight]{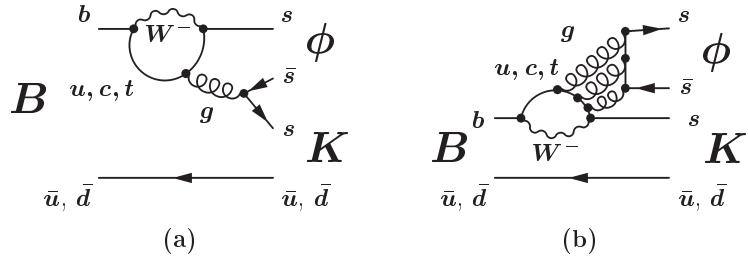}
  \caption{Example of quark level diagrams for $\B\to\phi K$}
  \label{fig:phiKs_diag}
\end{center}
\end{figure}

\begin{figure}
\begin{center}
        \includegraphics[width=0.47\textwidth]{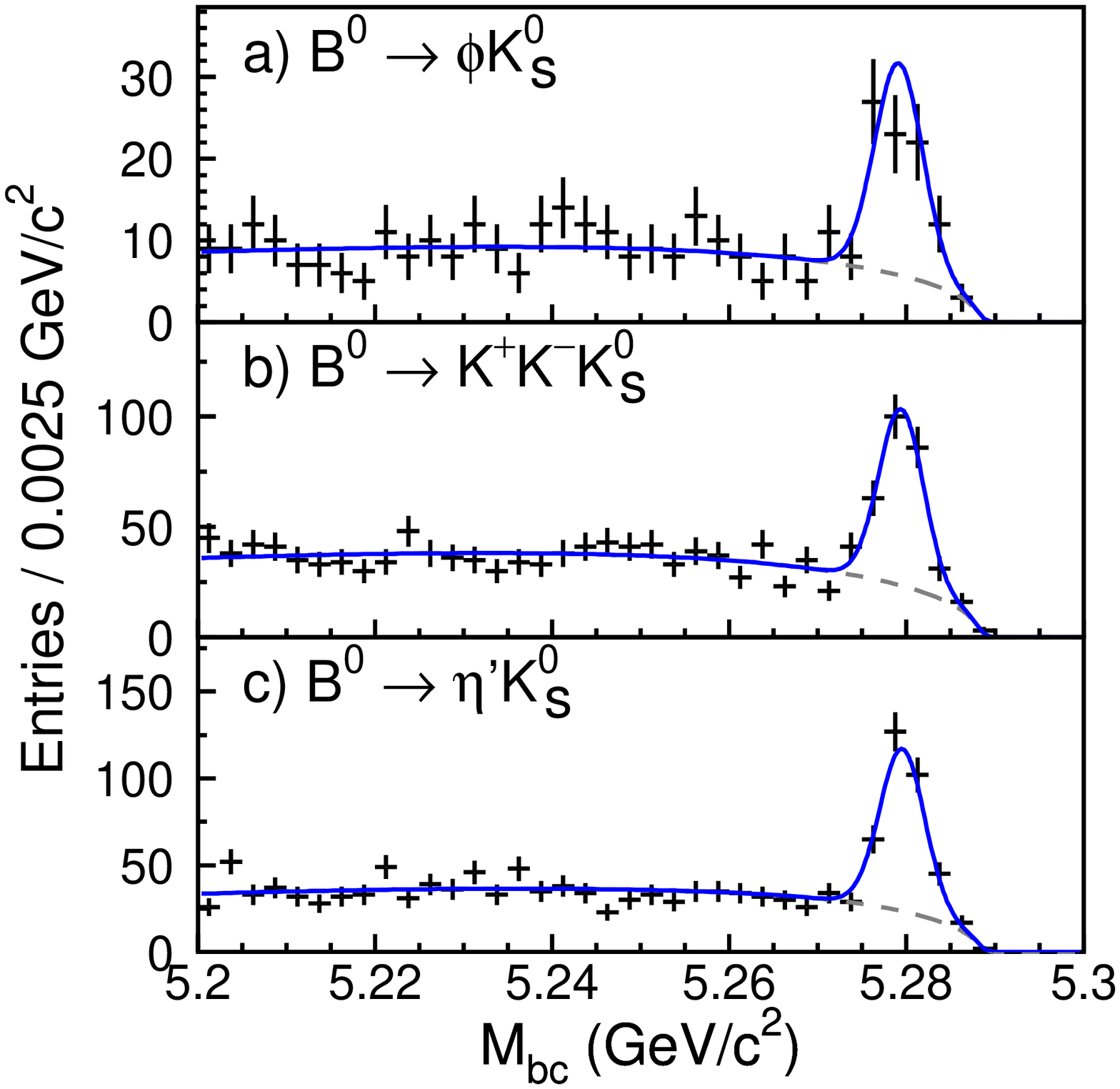}
\hfill
        \includegraphics[width=0.47\textwidth]{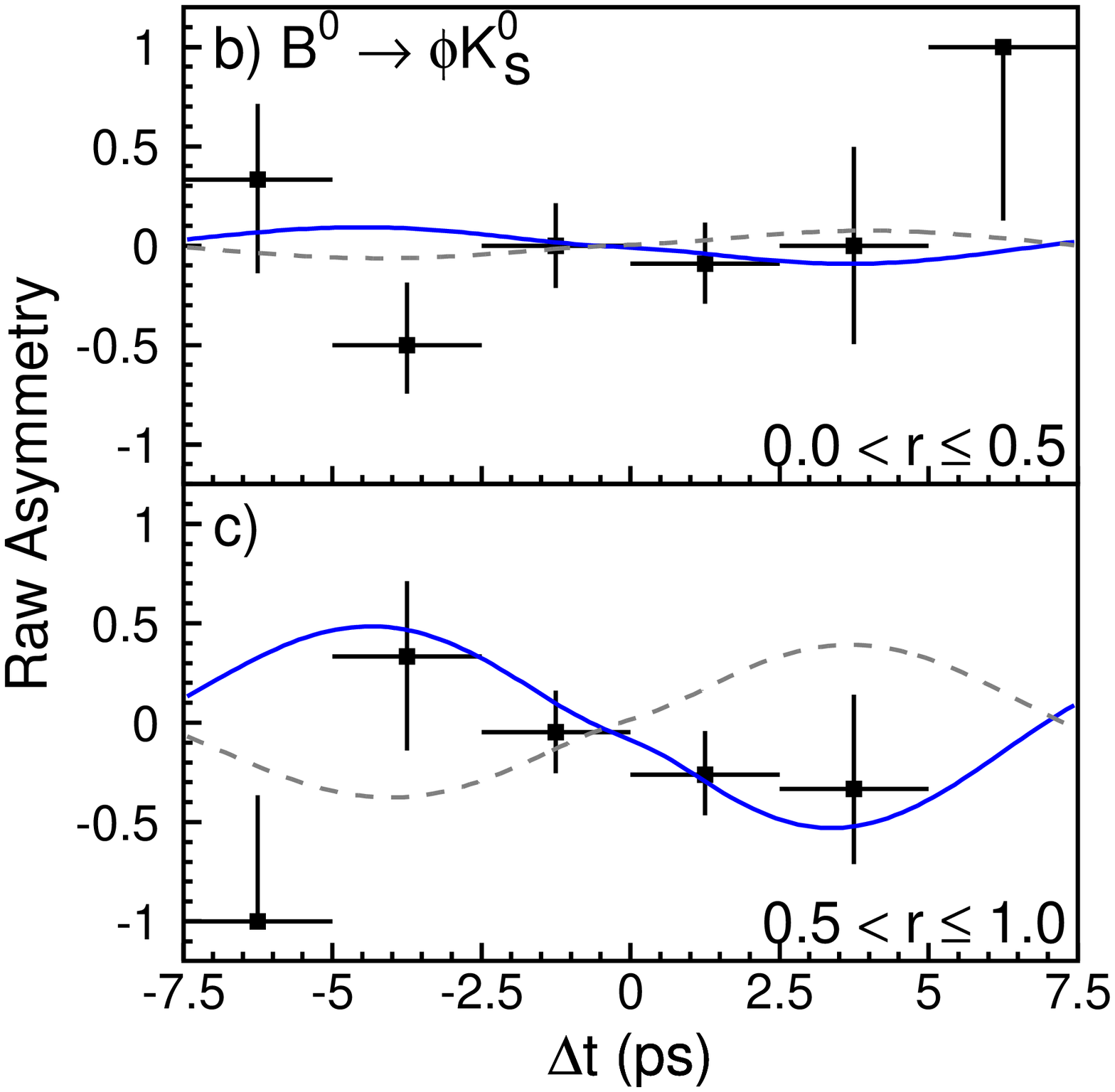}
\\
\caption{The beam-energy constrained mass distributions for three penguin dominated modes: $\phi\KS, K^+K^-\KS,\ \eta'\KS$ (left) and 
the raw asymmetry for $\phi\KS$ decay (right) mode measured by {\it BELLE}}
\label{fig:phiKs_Belle}
\end{center}
\end{figure}

In the SM decays like $\Bz \to \phi\KS$
are dominated by the $b\rightarrow s\bar{s}s$ gluonic penguin diagrams shown in Figure~\ref{fig:phiKs_diag}.
We expect $C=0$ in the SM because there is only one dominant decay mechanism.
Since $\phi\KS$ decays proceed through a \CP -odd  final state,
we expect $S=\sin{2\beta}$. The other contributions in the SM which can deviate the measured asymmetry from  
$\sin2\beta$ are rather small and range from several percents for $\phi\KS$ to some tens percents for others~\cite{ref:sPengCorr}.
However, contributions from physics beyond the Standard Model (NP), could invalidate these predictions~\cite{Grossman:1996ke}.
Since $b\rightarrow s\bar{s}s$ decays involve one-loop transitions, they are especially sensitive to such contributions.
Figure~\ref{fig:phiKs_Belle} shows the beam-energy constrained mass distributions for the three modes: $\phi\KS, K^+K^-\KS,\ \eta'\KS$, obtained by {\it BELLE}.
Clear peaks at the \B mass  demonstrate the ability to reconstruct modes with relatively small branching fractions of the order of $\sim 10^{-4}$~\cite{ref:phiKs_belle}. 

The {\it BELLE} \CP violation result obtained with about 152 M \BB pairs indicates a deviation from the $\sin2\beta$ value obtained with charmonium modes
of about 3.5$\sigma$:
\[
S_{\phi K^0}=-0.96\pm0.50(stat)^{+0.09}_{-0.11}(syst)
\]
Figure~\ref{fig:phiKs_Belle} also shows the raw asymmetry  for such a mode with the SM expectation overlaid.
The \babar\ results obtained with a similar data sample~\cite{ref:phiKs_babar}
\[
S_{\phi K^0}=+0.47\pm0.34(stat)^{+0.08}_{-0.06}(syst),
\]
is consistent with $\sin2\beta$.
In addition to $\phi\KS$ this result includes the  \CP asymmetry measured with  \CP-even $\phi\KL$ decay mode.
However, the two experiments  are in marginal agreement  within experimental errors for this 
decay.~\footnote{The recent results presented in~\cite{ref:ichep_plen} solves this problem. The two results are now in good agreement.}

A more accurate \CP\ violation measurement can be made using all decays to
$KK\KS$ that do not contain a $\phi$ meson.
This sample is several times larger than the sample of $\phi\KS$, but
the \CP\ content of the final state is not known.
The \CP content can be determined from isospin symmetry assumptions and measured branching fractions of $KK\KS$ and $K\KS\KS$
decays. Using this approach~\cite{ref:KKKs_belle} one observes that the \CP-even  state is strongly dominating decay channel
($f_{even}= 0.98\pm0.15\pm0.04$).
It is fortunate because it maximizes the experimental sensitivity on \CP violation.
Two results reported in~\cite{ref:phiKs_belle,ref:KKKs_babar} 
\begin{eqnarray*}
- S_{KK \KS}&=&+0.51\pm0.26(stat)\pm0.05^{+0.18}_{-0.00}(syst)\ (BELLE) \\
- S_{KK \KS}&=&+0.57\pm0.26(stat)\pm0.04^{+0.17}_{-0.00}(syst)\ (\babar) 
\end{eqnarray*}
are in a good agreement with the SM expectation.

\begin{figure}
\begin{center}
  \includegraphics[height=.3\textheight]{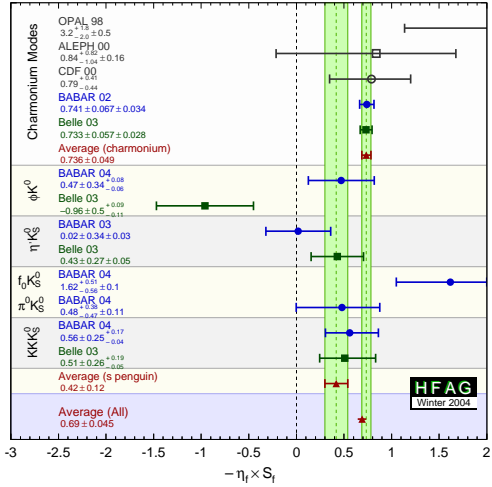}
  \caption{Compilation of the results for $-\eta_f\times S$  }
  \label{fig:sin2b_all}
\end{center}
\end{figure}

Figure~\ref{fig:sin2b_all} summarizes the measured \CP asymmetry relevant to $\sin2\beta$ for the charmonium and penguin dominated modes~\cite{ref:HFAG}. 
The $2.4\sigma$ difference in average between the two types of decays does not allow us to state whether it is or is not an effect of NP. 
It is important to continue this study to improve the experimental uncertainty until it is resolved.

\section{CKM phase {\boldmath $\alpha(\phi_2)$}}

\begin{figure}
\begin{center}
\begin{minipage}{0.48\textwidth}
  \includegraphics[width=0.8\textwidth]{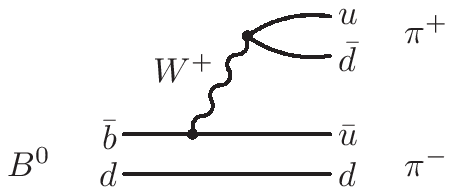}
\end{minipage}
\hfill
\begin{minipage}{0.48\textwidth}
  \includegraphics[width=0.8\textwidth]{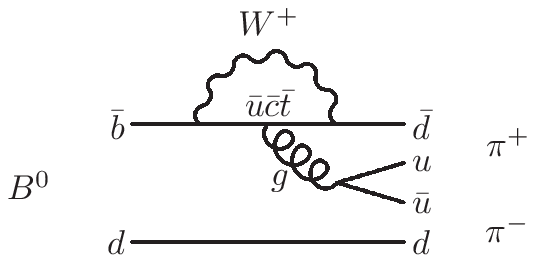}
\end{minipage}
  \caption{Tree (left) and gluonic penguin (right) diagrams contributing to the process $\B\to\pi\pi$}
  \label{fig:pipi_diag}
\end{center}
\end{figure}

In contrast to the theoretically clean measurements of $\sin2\beta$ with charmonium final states, 
the extraction of $\sin2\alpha$ is complicated by the presence of tree
and gluonic penguin amplitudes in modes like $\B\to hh$, where $h=\pi,\rho$.
Neutral $B$ decays to the \CP\ eigenstate
$\pip\pim$ can exhibit mixing-induced \CP\ violation through interference between
decays with and without $\Bz$--$\Bzb$ mixing, and direct \CP\ violation through
interference between the $b\to u$ tree and $b\to d$ penguin decay processes shown in Figure~\ref{fig:pipi_diag}.
Both effects are observable in the time evolution of the asymmetry between $\Bz$ and $\Bzb$
decays to $\pip\pim$, where the interference between decay and mixing  leads to a sine oscillation
with amplitude $S_{\pi\pi}$ and direct \CP\ violation leads to a cosine oscillation with
amplitude $C_{\pi\pi}$.  
In the absence of the penguin process, $C_{\pi\pi} = 0$ and $S_{\pi\pi} = \sin2\alpha$.
while significant tree-penguin interference leads to $C_{\pi\pi} \neq 0$ and
$S_{\pi\pi} = \sqrt{1 - C_{\pi\pi}^2}\sin{2\alpha_{\rm eff}}$.
The presence of loop (penguin) contributions introduces additional phases 
which can shift the experimentally measurable parameter $\alpha_{eff}$
away from the value of $\alpha$.
The difference between $\alpha_{\rm eff}$ and $\alpha$ can be determined from a
model-independent analysis using the isospin-related decays $B^{\pm}\to\pipm\piz$ and
$\Bz,\,\Bzb\to\piz\piz$~\cite{ref:GronauLondon}.

Results  on \CP violation in $\Bz,\,\Bzb\to\pip\pim$ decay mode are summarized in Table~\ref{tab:pipi} taken from Ref.\cite{ref:pipi_babar,ref:pipi_belle}. 
The {\it BELLE} experiment rule out the \CP-conserving case,
$S_{\pi\pi}=C_{\pi\pi}=0$ at the $5.2\sigma$ level. It also finds evidence of direct \CP violation with a  significance of $3.2\sigma$. The \babar\ collaboration does not confirm
the observation of large \CP violation in this decay channel reported by {\it BELLE}. However, the two results are in agreement within experimental errors.  

\begin{table}
\begin{center}
\begin{tabular}{ccc}  \hline
Parameter       &\babar\ (123 M \BB) & {\it BELLE} (152 M \BB) \\ \hline
$S_{\pi\pi}$    &$-0.40\pm0.22(stat)\pm0.03(syst) $&$-1.00\pm0.21(stat)\pm0.07(syst) $ \\
$C_{\pi\pi}$    &$-0.19\pm0.19(stat)\pm0.05(syst) $&$-0.58\pm0.15(stat)\pm0.07(syst) $ \\
$\rho(S,C)$                      & -0.02       & -0.29 \\ \hline
\end{tabular}
\label{tab:pipi}
\caption{Results on \CP violation measurements in $\Bz,\Bzb\to\pi^+\pi^-$. $\rho(S,C)$ is the correlation coefficient between $C$ and $S$ in the likelihood function.}
\end{center}
\end{table}

\begin{figure}
\begin{center}
\begin{minipage}{0.48\textwidth}
    \includegraphics[width=0.95\textwidth]{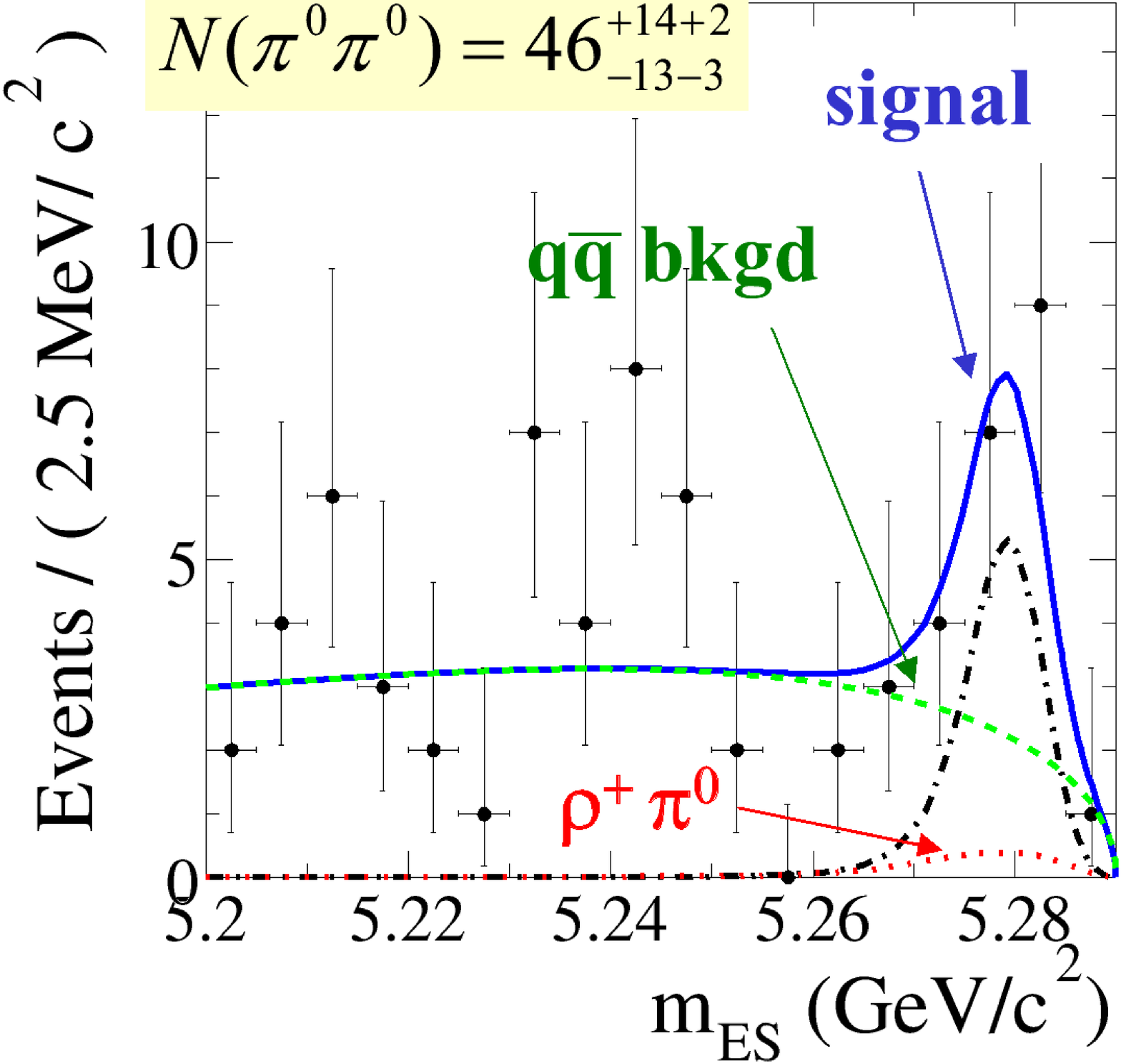} 
\end{minipage}
\hfill
\begin{minipage}{0.48\textwidth}
    \includegraphics[width=0.95\textwidth]{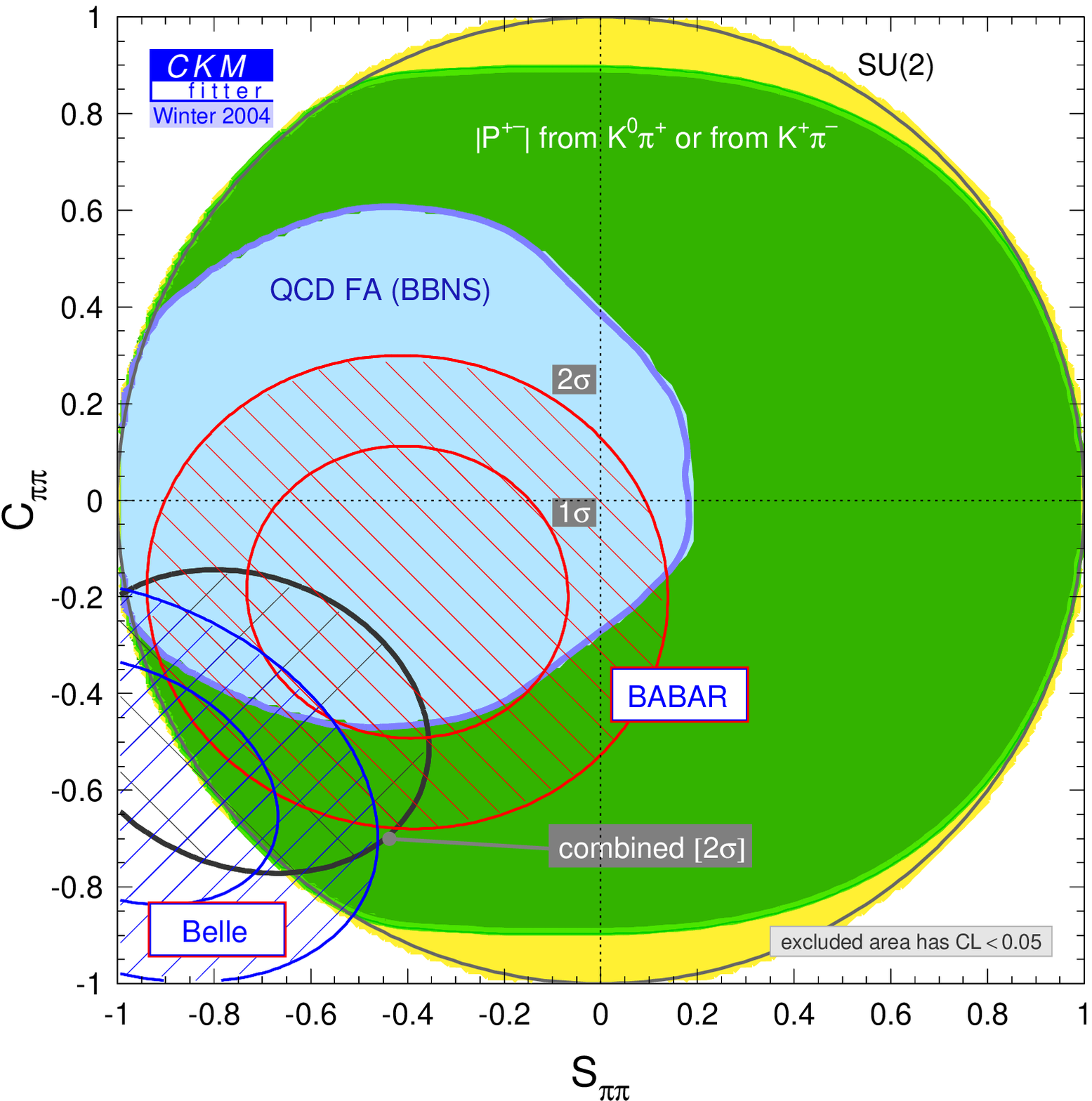}
\end{minipage}
\caption{The observation of $\Bz,\ \Bzb\to\pi^0\pi^0$ decay by \babar\ (left). 
The $1\sigma$ and $2\sigma$ contours for the \babar\ and {\it BELLE} in ($C$,$S$) plane obtained for  $\Bz,\Bzb\to\pi^+\pi^-$ decay (right).
}
\label{fig:pipi_SC}
\end{center}
\end{figure}

The difference between the measured $\alpha_{eff}$ and $\alpha$ is evaluated using measurements of the isospin-related decay $\Bz,\Bzb\to\pi^0\pi^0$. 
The observation of this decay, $4.2\sigma$ significance, by the \babar\ collaboration (Figure~\ref{fig:pipi_SC} (left))
with relatively large branching fraction~\cite{ref:pi0pi0_babar} demonstrates a large 
gluonic penguin contribution in this mode. However, this leads to essential difficulties for $\alpha$ extraction with $\B\to\pi\pi$ decays.

Figure~\ref{fig:pipi_SC} (right) shows a two-dimensional 68\% and 95\% C.L. for the experimental results in  the ($C$,$S$) plane. 
For comparison, the colored regions shows the 95\% C.L. obtained from the isospin analysis, the SU(3) $\Bp\to K^0\pi^+$ decay, and 
QCD factorization prediction. Large negative correlation between $S$ and $C$ observed in {\it BELLE} reflects the shape of the confidence region.
One can state that experimental results are consistent with isospin symmetry prediction, where knowledge of $\BR(\Bz\to\pi^0\pi^0)$
is still a dominant uncertainty.

The measurement of the $\B^\pm\to\rho^\pm\rho^0$ branching fraction and the upper limit for $\B^0\to\rho^0\rho^0$~\cite{ref:rho0rho0} indicate small
penguin contribution to the $B\to\rho\rho$ decay. Higher branching fraction and smaller shift of the measured parameters $\alpha_{eff}$ from 
$\alpha$ comparing to  $\Bz,\ \Bzb\to \pi^+\pi^-$ makes $\Bz,\ \Bzb\to\rho^+\rho^-$ decays more attractive for the extraction of the CKM angle $\alpha$. It is also fortunate
for the sensitivity to $\alpha$ that this two-vector final state is almost longitudinally polarized as it was measured in~\cite{ref:rhorho_pol} with an angular analysis.

\begin{figure}
\begin{center}
\begin{minipage}{0.48\textwidth}
    \includegraphics[width=0.9\textwidth]{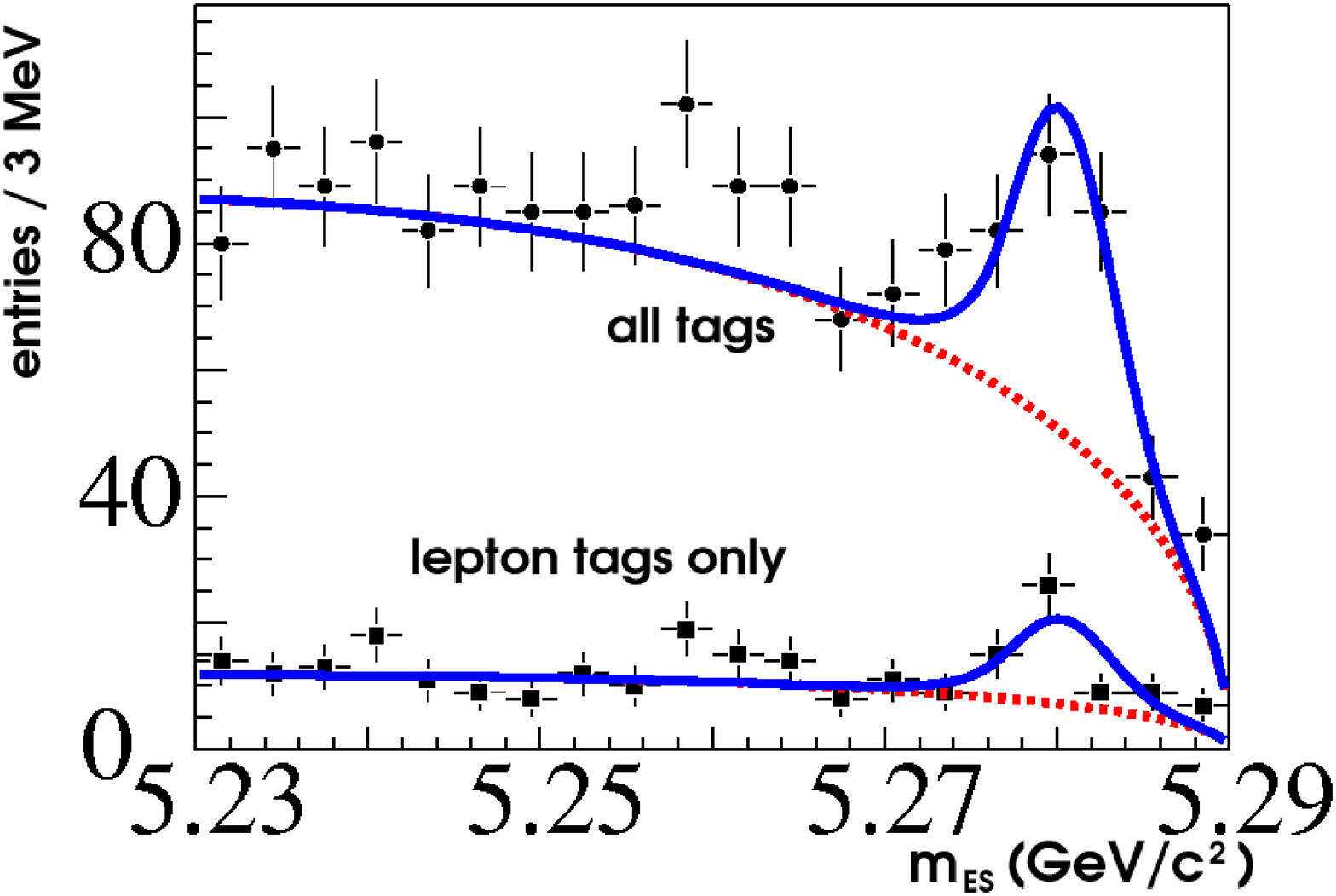}
\end{minipage}
\hfill
\begin{minipage}{0.48\textwidth}
    \includegraphics[width=0.9\textwidth]{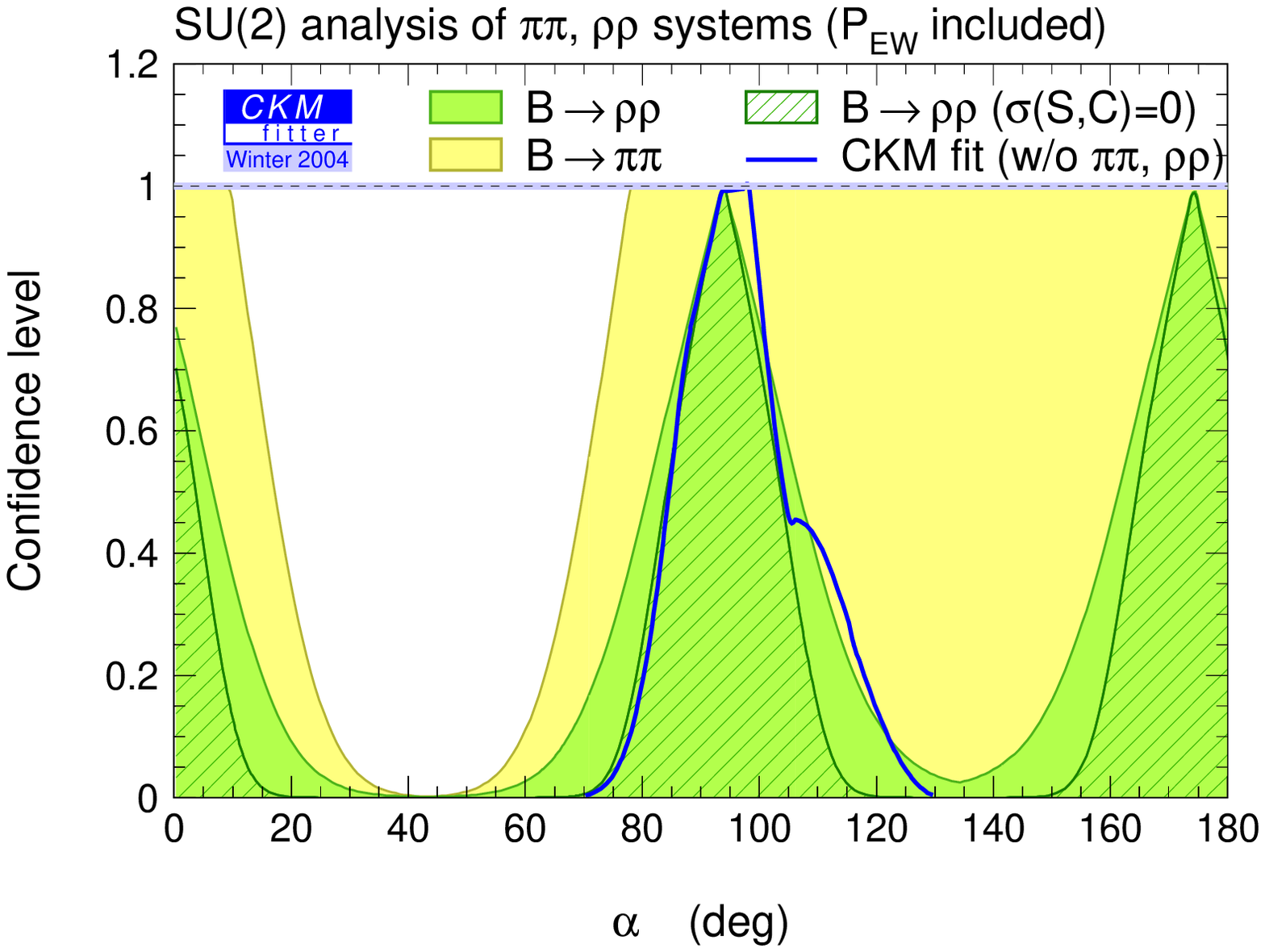}
\end{minipage}
\label{fig:rhorho}
\caption{The \B mass distribution for the $\Bz\to\rho^+\rho^-$ decay (left). Constrained on $\alpha$ obtained from the $\pi\pi$ and the $\rho\rho$ systems (right). The constraint 
assuming infinite precision for $C_{long}$ and $S_{long}$ is also shown. The plots are from \babar.}
\end{center}
\end{figure}

Figure~\ref{fig:rhorho} shows the \B mass distribution for the reconstructed $\rho^+\rho^-$ candidates~\cite{ref:rhorho_babar}. 
The \B candidates associated with only lepton tag, which provides the best signal-to-background ratio, are also shown. 
A clear peak at \Bz mass allows one to measure polarization and \CP asymmetry. 
The new \babar\ result for $\Bz,\ \Bzb\to\rho^+\rho^-$ decay, obtained with 123 million \BB pairs is the following:

\begin{eqnarray*}
     f_L &=& 1.00\pm0.02(stat)_{-0.03}^{+0.04}(syst) \\
      C_{long} &=& -0.23\pm 0.24(stat)\pm0.14(syst)  \\
      S_{long} &=& -0.19\pm 0.33(stat)\pm0.11(syst)
\end{eqnarray*}

Ignoring the possible non-resonant contributions, interference, I=1 amplitudes and assuming isospin symmetry, by using the experimental data on \BR($\Bz\to\rho^0\rho^0$), one can relate
the \CP parameters $S_{long}$ and $C_{long}$ to the CKM angle $\alpha$ up to a four-fold ambiguity. Selecting the solution closest to the CKM best 
fit average~\cite{ref:CKMFitter}, this corresponds to
\[
\alpha = 96^\circ \pm 10^\circ(stat)\pm 4^\circ(syst)\pm13^\circ(peng)
\]
where the last error is the additional contribution from penguins that is bounded at $<13^\circ$ (68.3\% C.L.)

Figure~\ref{fig:rhorho} (right) shows the constraint on $\alpha$ from the $\pi\pi$ and the $\rho\rho$ systems. \babar\ and {\it BELLE} 
average branching fractions, polarization in $\rho\rho$ (including the limit on $\rho^0\rho^0$, for which the polarization is unknown ) 
and asymmetry  $C$ and $S$ measurements are used to perform the Gronau-London isospin analysis. One can conclude that $\rho\rho$ system provides the most
precise constraint on $\alpha$, where the knowledge of penguin pollution is dominant.

\section{CKM phase {\boldmath $\gamma(\phi_3)$}}

Decays of $B_d$ mesons relevant to the CKM phase $\gamma$ show either small \CP asymmetry ($\B\to D^{(*)}\pi$) or branching fractions ($\B\to D^{(*)}K$). 
This produces essential difficulties for this measurement, where most of analyses are model dependent.
That is why  future experiments at Hadron Colliders are attractive for they will have access to the physics of the $B_s$ mesons.

\subsection{\boldmath \CP asymmetry with  $\Bz \rightarrow D^{(*)\mp} \pi^{\pm}$} 
The decay modes $\Bz \rightarrow D^{(*)\mp} \pi^{\pm}$ have been
proposed  to measure
$\sin(2\beta+\gamma)$~\cite{ref:sin2bg_th}.
In the Standard Model the decays
$\Bz \to D^{(*)+} \pi^-$ and $\Bzb \to D^{(*)+} \pi^-$
proceed through the $\overline{b} \rightarrow \overline{u}  c  d $ and
$\b\to c$ amplitudes $A_u$ and $A_c$, respectively.
The relative weak phase between these two amplitudes
is $\gamma$. When combined with $\Bz \Bzb$ mixing, this yields a weak phase
difference of $2\beta+\gamma$ between the interfering amplitudes.
The decay rate distribution for $B \to {D^{(*)}}^\pm\pi^\mp$ is described by 
an equation similar to (\ref{eq:timedist}), where
the parameters $C$ and $S$ are given by
\[
C \equiv {1 - \rsq \over 1 + \rsq}\, , \ \ \ \
S^\pm \equiv {2 \r \over 1 + \rsq}\, \sin(2 \beta + \gamma \pm \deltaPhase).
\]
Here $\deltaPhase$ is the strong phase difference
between $A_u$  and  $A_c$ and $r^{(*)} \equiv |A_u / A_c|$.
Since $A_u$ is doubly CKM-suppressed with respect
to $A_c$, one expects $r^{(*)}$ to be small of order 2\%.
Due to the small value of $r^{(*)}$, large data samples
are required for a statistically significant measurement of $S$. 
                                                                                                                                                                                  
Two different analysis techniques, full reconstruction
and partial reconstruction were used for 
the $\sin(2\beta+\gamma)$ measurement with $\Bz \rightarrow D^{(*)\mp} \pi^{\pm}$.

In the partial reconstruction of a $\btodstpipm$ candidate,
only the hard (high-momentum) pion track $\pi_h$ from the $B$ decay and the
soft (low-momentum) pion track $\pi_s$ from the decay
$D^{*-}\rightarrow \Dzb \pi_s^-$ are used.
Applying kinematic constraints consistent with the signal decay mode,
the four-momentum of the non-reconstructed, ``missing''
$D$ is calculated. Signal events are peaked
in the $\mmiss$ distribution at the nominal $\Dz$ mass.
This method eliminates the efficiency loss associated with the
neutral $D$ meson reconstruction. The \CP asymmetry independent 
on the assumption on $r^*$ measured with this technique by \babar is~\cite{ref:run1-2-ihbd} 
\[
2 r^{*} \sin(2\beta+\gamma) \cos \delta^{*} = -0.063\pm0.024\pm 0.014
\]                                                                                                                                        
This measurement deviates from zero by 2.3 standard deviations.
Both \babar\ and {\it BELLE} also use the full reconstruction technique ~\cite{ref:run1-2-breco,ref:s2bg_belle} to extract the $\sin(2\beta+\gamma)$value.

\begin{figure}
\begin{center}
\begin{minipage}{0.48\textwidth}
         \includegraphics[width=0.95\textwidth]{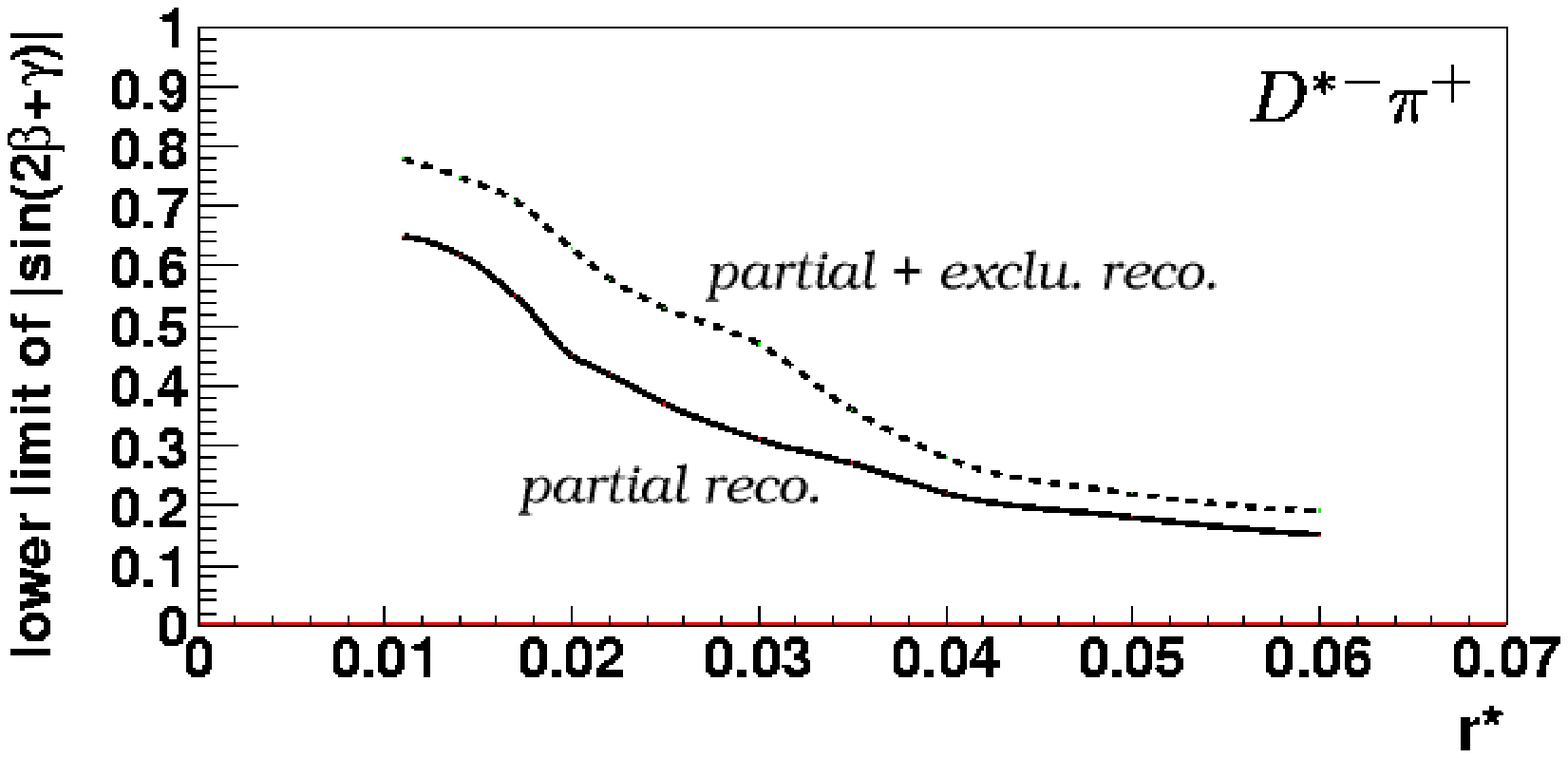}
\end{minipage}
\hfill
\begin{minipage}{0.48\textwidth}
    \includegraphics[width=0.95\textwidth]{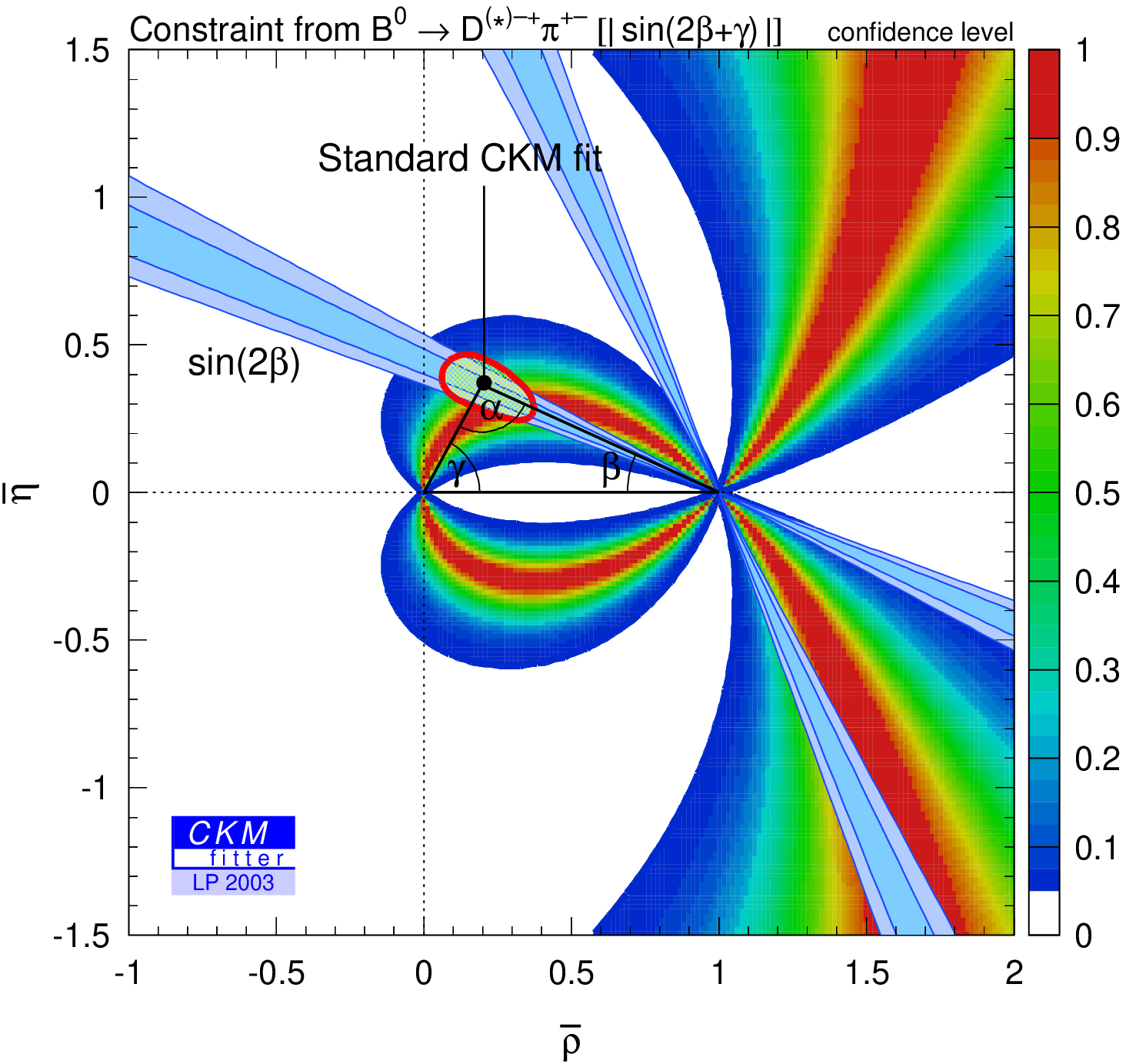}
\end{minipage}
\label{fig:sin2bg_constr}
\caption{95\% C.L. lower limit on $|\sin(2\beta+\gamma)|$ as a function of $r^*$ with \babar\ (left).
The solid curve corresponds to the partial reconstruction  analysis; the dashed curve includes the results of
full reconstruction for $\btodstpipm$ only. Probability contours (right) for the position of the apex of the unitary triangle
based on the $\Bz\to D^{(*)\mp}\pi^\pm$ decays.}
\end{center}
\end{figure}

Two methods for interpreting these results in terms of
constraints on $|\sin(2\beta+\gamma)|$ are used. 
Both methods involve minimizing a $\chi^2$
function that is symmetric under the exchange $\sin(2\beta+\gamma) \rightarrow
-\sin(2\beta+\gamma)$, and applying the method of Ref.~\cite{ref:Feldman}.
In the first interpretation method, no assumption regarding the value of $r^*$ is made.
The resulting 95\% lower limit for the mode $\Bz\to D^{*\mp}\pi^\pm$ is shown as a function of $r^*$ in
Figure~\ref{fig:sin2bg_constr} (left). 
The second interpretation assumes that $r^{(*)}$ can be estimated from the
Cabibbo angle, the ratio of branching fractions ${\cal
B}(B^0\rightarrow {D^{(*)}}_s^{+} \pi^-) / {\cal B}(B^0\rightarrow
{D^{(*)}}^{-} \pi^+)$, and the ratio of decay constants
$f_{\Dstar} / f_{\Dstar_s}$.
This method yields the limits~\cite{ref:run1-2-ihbd} 
$|\sin(2\beta+\gamma)|> 0.87$ at 68\% C.L. and
$|\sin(2\beta+\gamma)|> 0.58$ at 95\% C.L.
$|\sin(2\beta+\gamma)| = 0$ is excluded at 99.4 \% C.L.

\subsection{\boldmath $\gamma$ extraction  with $\B \rightarrow D^{(*)}K$ }

Several proposed methods for measuring $\gamma$ exploit the interference between $\Bm\to D^0K^-$ and $\Bm\to \bar{D}^0K^-$,
which occurs when $D^0$ and $\bar{D^0}$ decay into the same final state f. These methods are the following:
\begin{description}
\item[{\boldmath f$=K^+\pi^-$}] -
  CKM-suppressed (DCS) for \Dz and Cabibbo favored for \Dzb (ADS)~\cite{ref:ADS};
\item[{\boldmath f$=\pi^+\pi^-,\ K^+K^-,\ \KS\pi^0$}] - 
  \CP eigenstate (GLW)~\cite{ref:GLW}; 
\item[{\boldmath f$=\KS\pi^+\pi^-$}] - 3-body Dalitz plot analysis~\cite{ref:Dalitz}.
\end{description}
Theoretically clean measurements of the angle $\gamma$ can be obtained with ADS and GLW methods, while the Dalitz plot analysis relies on 
$D^0$ decay model. 

\begin{figure}
\begin{center}
    \includegraphics[width=0.5\textwidth]{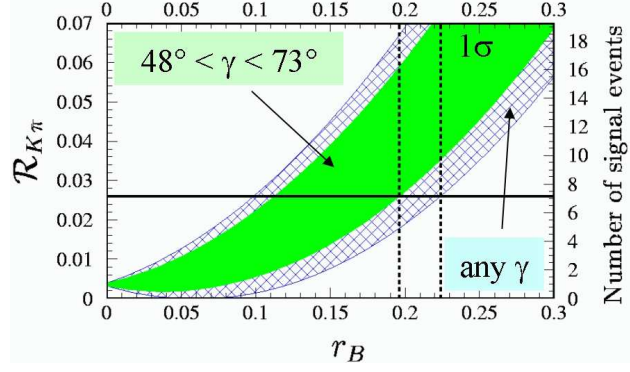}
\caption{Expectation on $R_{K\pi}$ and $N_{sig}$ versus $r_B$ obtained with \babar~\cite{ref:ads_babar}. }
\label{fig:ADS}
\end{center}
\end{figure}

The ADS method allows us to determine how large the suppression of $b\to u$  amplitude is.
Assuming no \CP violation in $D$ meson decays,  the measured quantity 

\[
R_{K\pi} = \frac{1}{2}R_{K\pi}^{+}+R_{K\pi}^{-}=r_B^2+r_D^2 + 2 r_B r_D\cos\gamma\cos (\delta_B+\delta_D), \
R_{K\pi}^{\pm} \equiv \frac{\Gamma([K^\mp\pi^\pm]_D K^\pm)}{\Gamma([K^\pm\pi^\mp]_D K^\pm)}
\]
where $r_B=\equiv \frac{|A(\Bm\to\Dzb\Km)|}{|A(\Bm\to\Dz\Km)|}\simeq 0.2$, $r_D \equiv |\frac{A(\Dz\to\Kp\pim)}{A(\Dz\to\Km\pip)}|=0.060\pm0.003$ can be used
to constraint $\gamma$. The analysis performed with 123 million \BB pairs yields $N_{sig.} = 1.1\pm3.0$ signal ($B^+\to [\Km\pip]_{D}\Kp$) candidates~\cite{ref:ads_babar}.
This allows one to calculate the Bayesian limit $R_{K\pi} <0.026$ at 90\% C.L. assuming a constant prior for $R_{K\pi}>0$. 
Figure~\ref{fig:ADS} shows the dependence of $R_{K\pi}$ on $r_B$. The area indicates the allowed region for any value of 
$\delta$, with a $\pm1\sigma$ variation on $r_D$ and the restriction with (filled-in) and without (hatched)  $48^\circ<\gamma<73^\circ$ constraint suggested by global 
CKM fit~\cite{ref:CKMFitter}. The 90\% C.L. upper limit on $r_B<0.196(0.224)$ with (without) the constraint on $\gamma$.  
To conclude, the small value of $r_B$, as suggested by this analysis,  makes determining  $\gamma$ from $\B\to DK$ difficult.

\CP-odd ($D^0\to\pi^+\pi^-,K^+K^-$)~\cite{ref:glw_babar,ref:glw_belle} and \CP-even ($D^0\to\pi^0\KS$, $\phi\KS$, $\omega\KS$, $\eta\KS$, $\eta'\KS$)~\cite{ref:glw_belle} 
decay modes were used to reconstruct $B^-\to D_{CP}^0K^-$. At the current precision of such measurements, $\gamma$ can not be constrained yet.

The CKM phase $\gamma$, $r_B$ and strong phase difference $\delta$ between the two amplitudes 
can be fitted in the Dalitz plot of $B^+\to [\KS\pim\pip]_{D}\Kp$. The decay model for Cabibbo allowed 3-body decay  of $D^0$ is measured in 
$D^*$-tagged $D^0$ decays. By using 152 million \BB pairs {\it BELLE} finds $35^\circ<\gamma<127^\circ,$  at 95\% C.L.~\cite{ref:dalitz_belle}. 
The fitted $r_B=0.31\pm0.11$ is somewhat large, but in agreement with ADS method.

\section*{Conclusion}
In conclusion, the two \B-factories have been operating successfully since 1999 and the \babar\ and {\it BELLE} experiments have already 
produced a lot of results relevant to the CKM phase measurements. Presence of \CP violation is well established in the \B-sector and 
its magnitude is in agreement with the CKM interpretation of this phenomenon in the Standard Model.
Measurements of the three CKM angles provide very important constraints on the apex of the Unitary Triangle 
There are several ``hot''  modes such as $\Bz\to \phi\KS$ (penguin dominated) and  $\Bz\to \pi^+\pi^-$ (presence of large penguin contribution), 
where statistical room for new effects exists.

\end{document}